\documentstyle[aps,prl,epsfig,epsf,multicol]{revtex}
\newcommand{\figwidth}{7cm}

\newcommand{\Am}{A_r}
\newcommand{\lp}{L_{\rm pin}}

\begin{document}

\title{Transient Dynamics of Pinning} 

\author{G. K. Leaf,$^1$ S.  Obukhov,$^2$ S. Scheidl,$^3$ and V. M.
  Vinokur$^4$}

\address{$^1$Mathematics and Computer Science Division, Argonne National
  Laboratory, Argonne, IL 60439
  \\
  $^2$University of Florida, Gainesville, FL 32611
  \\
  $^3$Institut f\"ur Theoretische Physik, Universit\"at zu K\"oln,
  Z\"ulpicher Str. 77, D-50937 K\"oln, Germany
  \\
  $^4$Materials Science Division, Argonne National Laboratory,
  Argonne, IL 60439}

\date{\today}
\maketitle

\begin{abstract}
  We study the evolution of an elastic string into the pinned state at
  driving forces slightly below the depinning threshold force $F_c$.
  We quantify the temporal evolution of the string by an {\it activity
    function} $A(t)$ representing the fraction of active nodes at time
  $t$ and find three distinct dynamic regimes.  There is an initial
  stage of fast decay of the activity; in the second, intermediate,
  regime, an exponential decay of activity is observed; and,
  eventually, the fast collapse of the string towards its final pinned
  state results in an decay in the activity with $\Am \sim
  (t_p-t)^{\psi}$, where $t_p$ is the pinning time in the finite
  system involved.
\end{abstract}


\begin{multicols}{2}
\narrowtext

Driven dynamics of disordered elastic media, which has become a
paradigm for a diversity of physical systems, is characterized by the
existence of a depinning force $F_c$, which separates the pinned state
at ``sub-threshold'' forces $F<F_c$ and the sliding regime at
``super-threshold'' forces $F>F_c$.  At finite temperatures the system
can move at sub-threshold forces as a result of thermally activated
jumps between the disorder-induced meta-stable states.  The
corresponding activation barriers diverge at small forces, $U(F) \to
\infty$ as $F \to 0$, giving rise to glassy dynamics with a highly
nonlinear response to infinitesimal drives \cite{IV,review,vm}
(creep), aging, memory, and hysteretic effects
\cite{fendr,andr,gord,metl}.  On the qualitative phenomenological
level, the observed memory and hysteretic behaviors can be viewed as a
result of slow relaxation of the relevant activation barriers upon
altering the drive\cite{stefan}, yet the detailed quantitative
description of transient glassy dynamics remains an appealing task.
The static sub-threshold characteristics of the string were studied in
\cite{mf} and critical depinning and transient behavior in the sliding
regime in \cite{les}.  In this letter we investigate numerically and
analytically the transient sub-threshold zero-temperature dynamics of
an elastic string, taken as an exemplary system for a driven elastic
disordered medium.  We study the time evolution of the string as it
slows and eventually stops upon applying a constant driving force in
the vicinity of the depinning threshold, $F \lesssim F_c$, where the
transient period and the distance traveled by the string are large and
diverge as $F \to F_c$.

\begin{figure}
  \epsfig{file=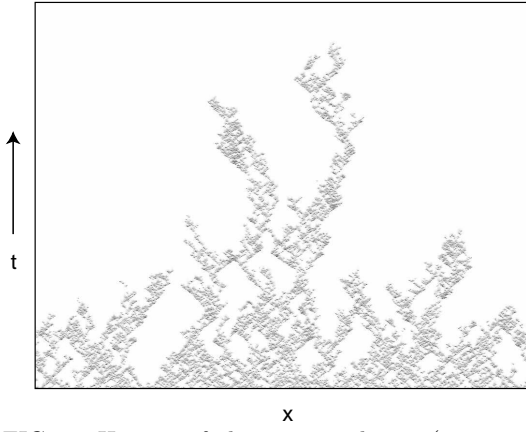,width=\figwidth} 
\vfill
  \caption{History of the string velocity (gray-scale plot of $v(x,t)$ using 
    white for zero velocity and black for large velocity). Note that
    in the active regions (avalanches) the velocity of the beads
    reaches the same values at early and late stages of the evolution,
    which differ only in the number and size of avalanches, which is
    measured by the activity.}
\label{fig.history}
\end{figure}

The graphic representation of a typical time evolution of the string
velocity $v(x,t)$ is given in Fig. \ref{fig.history}.  It demonstrates
the avalanche-like sub-threshold dynamics of the string.  The
avalanches advance via the propagation of kinks along the string.  We
quantify the temporal evolution by the activity function $A(t)$
defined as the fraction of the moving nodes at time $t$, which is
proportional to the velocity of the center of mass of the string
\cite{nodes}, and find three distinct regimes of the temporal
evolution of the string: (i) an initial regime of a fast decay of
activity, according to a power law; (ii) the exponential decay of
activity, $A \sim \exp\left[-(t/t_{\Delta})^{\gamma}\right]$ and (iii)
the final avalanche-like immobilization stage, which appears only for
strings of finite length for which the activity vanishes and the
string gets pinned at a finite time $t_p$.  Shortly before $t_p$ the
activity function drops rapidly to zero following the power law, $\Am
\sim (t_p-t)^{\psi}$, with $\psi\approx 0.63$.

{\it Model --} We consider an elastic string at zero temperature
placed on a disordered plane and subject to a constant drive $F$ not
exceeding the critical depinning force $F_{c}$.  The overdamped
equation of motion is $\eta {\partial y}/{\partial t} = -{\delta {\cal
    H}}/{\delta y}$, where $y(x,t)$ is its lateral displacement at
time $t$, $\eta$ is a friction coefficient.  The string has a length
$L$ and is aligned along the $x$ direction.  Its energy is
\begin{eqnarray*}
{\cal H} = \int_0^L dx \bigg[\frac{C}{2}\left(\frac{\partial
y}{\partial x}\right)^2 + U(y(x),x) \bigg] ,
\end{eqnarray*}
with the string elastic constant $C$. The string and the media are
periodic in $x$ with periodicity $L$.

In the simulations we use a discrete version of the continuum
equations.  The $x$ direction is discretized by a set of lines (rails)
of equal spacing, with unit length, constraining the motion of
discrete string elements (nodes).  Energies are measured in the units
of the elastic constant $C$.  The  pinning sites are
randomly distributed on the rails according to a specified density
$\rho = 0.1$. The pinning sites are triangular wells with unit width
and the depth/strength $U_p=0.1$.  Forward Euler time marching is used
to advance the system in time.  The initial state for the string is
taken as a straight line.  The statistics were collected on $30$
samples of length $L=8192$.

\begin{figure}
  \epsfig{file=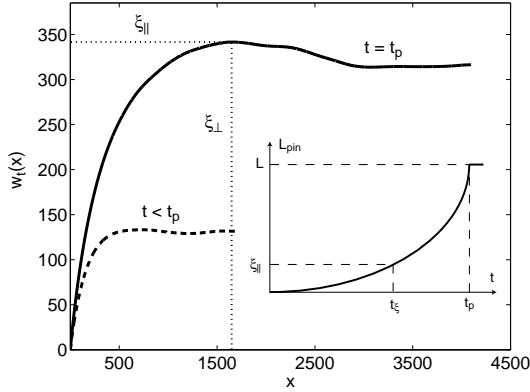,width=\figwidth} 
\vfill
  \caption{Plot of the roughness $w_t(x)$ at intermediate time
    ($t<t_p$, truncated for large $x$ beyond saturation) at final
    stage ($t\geq t_p$), where it saturates on the scale of the
    correlation lengths $\xi_{\parallel, \perp}$. The nonmonotonic
    shape of $w_t(x)$ is an artifact of the finite number of samples
    used for disorder averaging. Inset: time evolution of the typical
    size $\lp$ of pinned string segments.}
  \label{fig.w}
\end{figure}

\begin{figure}
  \epsfig{file=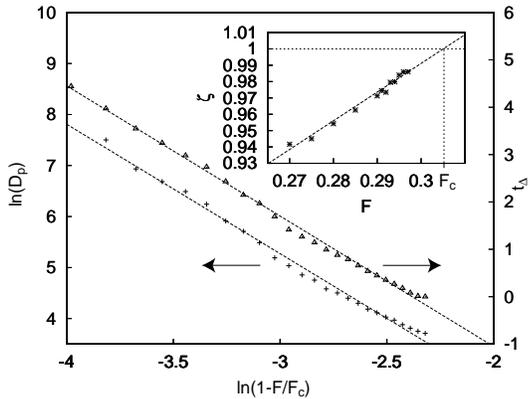,width=\figwidth} 
\vfill
  \caption{Inset: The roughening  exponent $\zeta$ as a function of the 
    driving force $F$ (stars: measured values, dashed line: linear
    fit), which assumes the value $\zeta=1$ at the critical force
    $F=F_c$.  Main plot: force dependence of the pinning distance
    $D_p$ (crosses) and of the relaxation time $t_\Delta$ (triangles)
    with linear fits (dashed lines) in the double-logarithmic
    representation.}
\label{fig.F}
\end{figure}

{\it Results --} To link to past research, we measured the geometric
characteristics of the subthreshold behavior.  The roughness of the
string at time $t$ is $ w_t(x) = \langle [y(x+x',t) -y(x',t)]^2
\rangle^{1/2}$, where $\langle \dots \rangle$ denotes a positional
average over $x'$ and a disorder average.  The roughness increases
with $x$ (for $x \ll L$) until it saturates at $x \sim
\xi_\parallel(t)$ with $w_t(x)\approx \xi_\perp(t)$ (see Fig.
\ref{fig.w}).  Correlation lengths increase with time and reach the
final values $\xi_\parallel$ and $\xi_\perp$ in the pinned state ($t
\geq t_p$), where the roughness follows a power law $w_\infty \sim
x^{\zeta}$ for $1 \ll x \ll \xi_\parallel$ \cite{hh}.  We find the
roughening exponent $\zeta$ to grow linearly as the applied force
approaches its threshold value $F_c$ (see inset of Fig.  \ref{fig.F}).
This behavior allows for an alternative determination of $F_c$ resting
on the analytical estimate $\zeta=1$ at the transition \cite{nar}.
This contrasts with the usual procedure where two parameters $F_c$ and
$\zeta$ have to be simultaneously determined as fitting parameters.
The criterion $\zeta(F_c)=1$ yields as an estimate for the critical
depinning force $F_c = 0.3016$, which is in excellent agreement with
our value of the critical force determined independently from the
double-logarithmic plot of roughness ($F_{c} = 0.3105$) and also with
the earlier finding \cite{dong}($F_c = 0.3058$).  Note a discrepancy
with the results in Ref.  \cite{les}, where $\zeta =
1.25$ at the depinning transition has been found in conflict with the
analytical result of \cite{nar}.  The reason for this discrepancy is
not yet clear.

Next, we studied the behavior of the pinning distance \cite{mf} $D_p$,
which is defined as the ensemble averaged distance that the center of
mass of the string traveled before being completely pinned
$D_p=(1/L)\sum_{i=1}^L|y_i(t_p)- y_i(0)|$.  We verified the scaling
behavior of $D_p \sim (F-F_c)^{-\alpha}$ (cf. Fig.  \ref{fig.F}). For
the given string length $L = 8192$ the finite-size effects become
negligible for the force dependence.  We found $\alpha=2.56\pm 0.08$
for $F_c=0.3016$ in good agreement with the results of \cite{mf} for
the 1D CDW.  The other relevant quantities $\xi_\parallel$ and
$\xi_\perp$, determined from the saturation of the roughness also
scale with exponents grouping around $\alpha \approx 2.4$.  The
results are summarized in the Table \ref{tab}, which also gives the
values of the critical force determined independently from the
divergences of the quantities involved.

\begin{figure}
  \epsfig{file=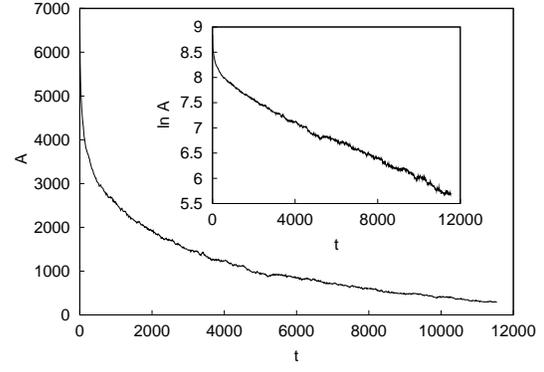,width=\figwidth} 
\vfill
  \caption{Time dependence of the activity $A$ for the force 
    value $F=0.29$ in linear and semilogarithmic representation.}
\label{fig.A.t}
\end{figure}

Having established the geometric characteristics of subthreshold
dynamics, we now turn to a detailed study of the temporal evolution of
transient behavior.  We characterize the temporal evolution of a
finite system toward the pinned state by the {\it activity function}
$A(t)$ defined as the fraction of the moving string nodes moving at
the given instant $t$. For a discrete system evolving in time by steps
$\delta{t}$, we define a node as active at time $t$ if that node has
moved at least one lattice unit in the previous 100 time steps. The
activity function is then the ensemble average of the fraction of
active nodes. In Fig.  \ref{fig.A.t} we illustrate the temporal
behavior of the activity function for the representative subthreshold
force $F = 0.29$.

It is important to quantify not only the fraction of active nodes, but
also the spatial structure of pinned and moving nodes, which can be
quantified by the characteristic length $\lp$ of pinned segments
(connected region of pinned nodes).  Initially, $\lp$ will be of the
order of unity, since the string position is not correlated with the
location of the pinning centers and in the final pinned state $\lp =L$
(see inset of Fig. \ref{fig.w}).  The initial transient regime (i) is
expected to last until $t_\xi$ given by $\lp(t_\xi)=\xi_\parallel$
\cite{les}.

At $t>t_\xi$ relatively large segments of the string are pinned, and
the remaining activity is due to the motion of mobile segments over
rare regions, where the pinning centers are underrepresented.  These
mobile segments will advance until the driving force is balanced by
other pinning centers or by the string tension.  We now construct a
lower estimate for the velocity of an {\em infinitely} long string as
a function of time.  We expect the finite string in regime (ii) to
follow this dynamics, before finite-size effects set in in regime
(iii).  We divide the rails in elements of unit length.  Then a finite
fraction $p$ of elements is free of a pinning force.  We overestimate
the effect of pinning by assuming that the other elements contain an
infinitely strong pinning force.  Let us now consider a string segment
that is pinned on two rails of distance $l$, say at $y(0)=y(l)=0$.  If
the rails in between are free of pinning centers, the segment relaxes
into the configuration where the driving force is balanced by the
string tension.  A simple estimate shows that the front of the segment
will then reach a distance $y_l \sim F l^2/C$ after a time $t_l \sim
y_l \eta/F \sim \eta l^2/C$.  Before this time the ends of length
$x(t) \sim \sqrt{Ct/\eta}$ of the string segment come to rest, and the
center of mass of the segment moves with a velocity $v_l(t) \sim
(F/\eta)[1-\sqrt{Ct/\eta l^2}]$.  During this relaxation the string
segment moves over an area ${\cal A}_l \sim F l^3/C$. The relaxation
takes place in this undisturbed way only if all the rail segments in
this area are free of pinning, which occurs with probability $p_l \sim
p^{{\cal A}_l}$.

To find a lower bound to the string velocity we divide the whole
string into segments of arbitrary length $l$, which serves as
variational parameter.  Although the string can be taken as unpinned
initially, the velocity is underestimated by formally considering the
links between these sections as pinned.  A fraction $p_l$ of such
segments relaxes without encountering pinning rail elements and
contributes $\sim p_l v_l(t)$ to the average string velocity. This
contribution is a lower bound to the velocity for all possible $l>
\sqrt{Ct/\eta}$, and the best estimate is obtained by $v_{\rm min}(t)
\sim \max_l p_l v_l(t) \sim p^{(F/C)(Ct/\eta)^{3/2}}$, which is finite
at all times.  Consequently, the activity function is also expected to
decay not faster than $A(t)\gtrsim \exp(-{\rm const.} t^{3/2})$.

In the {\em finite} system this exponential decay stops when the
characteristic distance between the moving segments $\lp \sim l/p_l$
becomes of the order of the system size and the string experiences a
deficiency of the moving nodes.  Our conjecture is that the origin of
this ``superfast'' stopping regime is the avalanche-like disappearance
of the moving nodes of the string. Once pinned, a node cannot depin
any more; moreover, it causes the immediate pinning of the neighboring
nodes, in a close analogy with the avalanche clustering at the final
stage of the coagulation process \cite{poly}.  This analogy suggests
the power-law termination of the motion according to $\Am(t_p-t) \sim
|t_p-t|^\psi$, where $t_p$ is the time of the total immobilization.
Although $t_p$ is subject to strong sample-to-sample fluctuations
because it is determined by fluctuations on the scale of the system
size, the final regime can be recognized only by averaging the
activity considered as a function of the reverse time $t_p-t$; that
is, the final regime can be seen only if the averaging over $N$
samples is performed according to $\Am(t_p-t) = (1/N) \sum_{i=1}^N
A_i(t_p^{(i)}-t)$ in contrast to the usual average $A(t)= (1/N)
\sum_{i=1}^N A_i(t)$.

We give lower and upper limits on the exponent $\psi$.  Consider the
process in which the node that came to rest enhances the probability
that the nearest node will stop at the next moment of time. In this
case the boundary between localized and mobile nodes will move some
average velocity, and the activity will decrease as $\Am \propto
t_p-t$, which means that the upper limit on $\psi$ is $\psi_{\rm
  upper}=1$. This type of behavior can be expected when $F$ is far
below $F_{c}$.  Another limiting type of terminal dynamics occurs if
immobile nodes do not affect dynamics in mobile regions.  In this case
the boundary between the immobile and mobile regions moves randomly,
and the finite string has a nonzero chance of stopping when the
boundaries enclosing the moving region collide with each other.
Considering this process in reverse time, we conclude that $\Am
\propto (t_p-t)^{1/2}$, and we get a lower limit $\psi_{\rm
  lower}=1/2$.  Note, that the possibility of fragmentation of the
moving segment of the string into several subsegments does not affect
this estimate.  In the more general case of a moving manifold of
internal dimension $d$ ($d=1$ for the string), similar arguments
result in $d/2 < \psi < d$.

\begin{figure}
  \epsfig{file=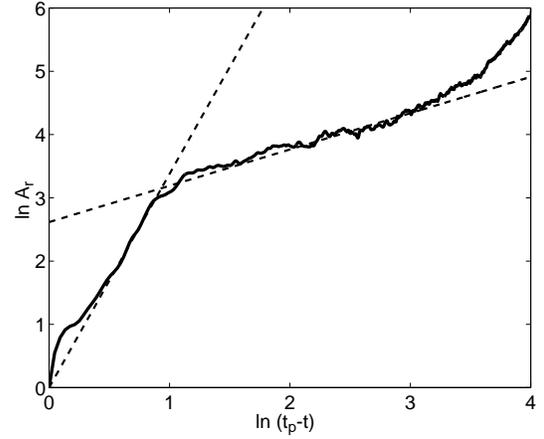,width=\figwidth}
\vfill
  \caption{Plot of the activity $\Am$ as a function of the reverse
    time $(t_p-t)$. The dashed lines represent linear fits to the
    regimes (ii) and (iii) introduced in the text.}
  \label{fig.final}
\end{figure}

Numerical simulations clearly show distinct relaxational regions.
After the initial relaxation (regime (i)), the activity crosses over
to the exponential decay regime, $A(t)\sim \exp
\left[-(t/t_{\Delta})^{\gamma}\right]$, $\gamma\simeq 1$. The
precision of our data is not sufficient to determine a more precise
estimate of $\gamma$; from the above consideration we could expect
$1<\gamma\leq 3/2$.  The relaxation time $ t_{\Delta}$ diverges as the
applied force approaches the threshold value (see Fig. \ref{fig.F}).
However, although one could expect that $ t_{\Delta}=t_{\xi}$ and
therefore that $t_{\Delta} \sim |F_c-F|^{\nu z}$ (since $\xi \sim
|F-F_c|^{\nu}$), with $z\nu=4/3$ using the exponents of the depinning
transition \cite{les}, we have found a different exponent: $t_{\Delta}
\sim |F_c-F|^{\Delta}$ with $\Delta = 2.74 $.  A careful examination
of a final stage of the relaxation process reveals a very sharp drop
of the activity function just before the complete immobilization of
the string.  To analyze this stage, we introduce the {\it reverse
  time} representation: we present the activity function as a function
of $t_r=t_p-t$, counting time backwards from the moment of achieving
final immobilization.  Using this representation enables us to carry
out an ensemble averaging despite the fact that different samples have
different pinning times.  The plot of $\ln \Am$ vs. $\ln(t_p-t)$
reveals a power-law final behavior (see Fig.  \ref{fig.final}): $\Am
\sim (t_p-t)^{\psi}$, with $\psi = 0.63\pm 0.06$, which is within the
above upper and lower estimates. We note that the reverse time plot of
$\ln \Am$ reconfirms the existence of the intermediate regime with the
essentially same characteristic time $t_{\Delta}$ as was found from
the direct time picture.

In summary, we investigated subthreshold dynamics of a pinned elastic
string and identified three distinct transient regimes: (i) a fast
initial relaxation, (ii) an intermediate exponential decay of the
activity resulting from residual motion in the exponentially rare
regions free of defects and (iii) a novel avalanche-like terminal
relaxation to the pinned state, $\Am \sim (t_p-t)^{\psi}$, resulting
from the finite-sized effects.  This final stage exhibits striking
similarity to coagulation dynamics.

This work was supported by Argonne National Laboratory through the
U.S. Department of Energy, BES-Material Sciences and BES-MICS, under
contract No.  W-31-109-ENG-38, and by the NSF-Office of Science and
Technology Centers under contract No. DMR91-20000 Science and
Technology Center for Superconductivity.

\begin{table}
\begin{tabular}{|c|c|c|c|}
\hline
& $D_p$ &  $\xi_{\perp}$ & $\xi_{\parallel}$\\
\hline
$\alpha$ & $2.56\pm 0.08$ &  $2.41\pm 0.07$ & $2.38\pm 0.09$\\
\hline
$ F_{c}$ & $0.2978\pm 0.001$
& $0.3011\pm 0.002$ &$0.3007\pm 0.01$
\\
\hline
\end{tabular}
\vfill
\caption{Values of the scaling exponent $\alpha$ and the critical
  force $F_c$ as determined from the divergences of $D_p$, $\xi_\perp$,
  and $\xi_\parallel$.}
\label{tab}
\end{table}

\end{multicols}


\begin{references}

\bibitem{IV} L. B. Ioffe and V. M. Vinokur, J. Phys. C {\bf 20}, 6149
  (1987)

\bibitem{review} G. Blatter {\it et al.}, Rev. Mod. Phys. {\bf 66}, 1125
  (1994)

\bibitem{vm} V. M. Vinokur, M. C. Marchetti, and L.-W. Chen, 
Phys. Rev. Lett., {\bf 77}, 1845 (1996)

\bibitem{fendr} J. A. Fendrich {\it et al.},  \prl {\bf 77}, 2073 (1996).

\bibitem{andr} W. Henderson, E. Y. Andrei, and M. J. Higgins, \prl
 {\bf 81}, 2352 (1998)

\bibitem{gord} S. N. Gordeev {\it et al.}, Nature, {\bf 385}, 324 (1997)

\bibitem{metl} Metlushko {\em et al.}, unpublished

\bibitem{stefan} S. Scheidl and V. M. Vinokur, \prl {\bf 77}, 4768
  (1996).

\bibitem{mf} A. Middleton and D. Fisher, Phys. Rev. {\bf B 47}, 3530 (1993)
  
\bibitem{les} H. Leschhorn {\it et al.}, Ann. Phys. (Leipzig) {\bf 6},
  1 (1997).

\bibitem{nodes} In the subthreshold regime the motion of the string is
  avalanche-like.  Once a bead participates in an avalanche, it
  reaches a typical maximum velocity that is independent of the time,
  where the avalanche started.  The activity is defined by the
  fraction of particles with a velocity exceeding a threshold velocity
  smaller than the typical maximum velocity.  Therefore, the activity is
  proportional to the average velocity.

\bibitem{hh} T. Halpin-Healy and Y.-C. Zhang, Physics Reports {\bf
    254}, 215 (1995).

\bibitem{nar} O. Narayan and D. S. Fisher, Phys. Rev. {\bf B 48}, 7030
  (1993).
  
\bibitem{dong} M Dong {\it et al.}, Phys. Rev. Lett. {\bf 70}, 662
  (1993).
 
\bibitem{poly} M. H. Ernst, Kinetic theory of clustering, in {\it
    Fundamental problems of statistical mechanics} VI, Ed. E.G.D.
  Cohen, Elsevier Science Publ., B.V. (1985)



\end{references}
\end{document}